\pdfoutput=1

\documentclass[journal]{IEEEtran}
\ifCLASSINFOpdf
   \usepackage[pdftex]{graphicx}
   \graphicspath{{../pdf/}{../jpeg/}}
   \DeclareGraphicsExtensions{.pdf,.jpeg,.png}
\else
\fi
\usepackage{latexsym} 
\usepackage[cmex10]{amsmath}
\newcommand{\str}[1]{#1}	

\newcommand{\MYfooter}{\smash{\scriptsize
\hfil\parbox[t][\height][t]{\textwidth}{\centering
~\\
~}\hfil\hbox{}}}

\newcommand{\MYarxivheader}{\smash{\scriptsize
\hfil\parbox[t][\height][t]{\textwidth}{\centering
(c) 2014 IEEE. Personal use of this material is permitted. Permission from IEEE must be obtained for all other users, including reprinting/ republishing this material for advertising or promotional purposes, creating new collective works for resale or redistribution to servers or lists, or reuse of any copyrighted components of this work in other works.}\hfil\hbox{}}}

\makeatletter

\def\ps@headings{%
\def\@oddhead{\mbox{}\scriptsize\rightmark \hfil \thepage}
\def\@evenhead{\scriptsize\thepage \hfil \leftmark\mbox{}}
\def\@oddfoot{\MYfooter}%
\def\@evenfoot{\MYfooter}}

\def\ps@IEEEtitlepagestyle{%
\def\@oddhead{\MYarxivheader}%
\def\@evenhead{\scriptsize\thepage \hfil \leftmark\mbox{}}%
\def\@oddfoot{\MYfooter}%
\def\@evenfoot{\MYfooter}}

\makeatother

\begin{document}
%
\title{A Comprehensive Graphene FET Model for Circuit Design}
%
%
%

\author{Saul~Rodriguez,~\IEEEmembership{Member,~IEEE,}
	Sam~Vaziri,~\IEEEmembership{Student Member,~IEEE,}
        Anderson~Smith,~\IEEEmembership{Student Member,~IEEE,}
        S\'ebastien~Fr\'egon\`ese,        
	Mikael~Ostling,~\IEEEmembership{Fellow,~IEEE,}
        Max~C.~Lemme,~\IEEEmembership{Senior Member,~IEEE,}
       and~Ana~Rusu,~\IEEEmembership{Member,~IEEE}

\thanks{Manuscript received October 1, 2013; accepted January 21, 2014. Support from the European Commission through a STREP project (GRADE, No. 317839), an ERC Advanced Investigator Grant (OSIRIS, No. 228229), and an ERC Starting Grant (InteGraDe, No. 307311) as well as the German Research Foundation (DFG, LE 2440/1-1) is gratefully acknowledged.}
\thanks{S. Rodriguez, A. Smith, S. Vaziri, M. Ostling, and A. Rusu are with the KTH Royal Institute of Technology, School of ICT, Kista, Sweden (email: saul@kth.se; andsmi@kth.se; vaziri@kth.se; ostling@kth.se; arusu@kth.se)}
\thanks{S. Fr\'egon\`ese is with the CNRS and
Université de Bordeaux, France (email: sebastien.fregonese@ims-bordeaux.fr)}
\thanks{M. C. Lemme is with the University of Siegen, Graphene-based Nanotechnology, Germany (email: max.lemme@uni-siegen.de)}}
\maketitle


\begin{abstract}
During the last years, Graphene based Field Effect Transistors (GFET) have shown outstanding RF performance; therefore, they have attracted considerable attention from the electronic devices and circuits communities.  At the same time,  analytical models that predict the electrical characteristics of GFETs have evolved rapidly. These models, however, have a complexity level that can only be handled with the help of a circuit simulator. On the other hand, analog circuit designers require simple models that enable them to carry out fast hand-calculations, i.e., to create circuits using small-signal hybrid$\text{-}\pi$ models,  calculate figures of merit, estimate gains, pole-zero positions, etc. This paper presents a comprehensive GFET model that is simple enough for being used in hand-calculations during circuit design and at the same time it is accurate enough to capture the electrical characteristics of the devices in the operating regions of interest. Closed analytical expressions are provided for the drain current $I_D$, small-signal transconductance gain $g_m$, output resistance $r_o$, and parasitic capacitances $C_{gs}$ and $C_{gd}$. In addition, figures of merit such as intrinsic voltage gain $A_V$, transconductance efficiency $g_m/I_D$, and transit frequency $f_T$ are presented. The proposed model has been compared to a complete analytical model and also to measured data available in current literature. The results show that the proposed model follows closely to both the complete analytical model and the measured data; therefore, it can be successfully applied in the design of GFET analog circuits. 

\end{abstract}


%
\IEEEpeerreviewmaketitle


\section{Introduction}
%
%
%
%
\IEEEPARstart{T}{he} reduction of dimensions in Silicon based transistors faces great challenges as dimensions  approach atomic sizes and physical limits will be eventually reached.  A great deal of research has focused during the last years in new materials that alleviate these limitations. One of these materials is Graphene \cite{Novoselov2004}⁠, a two-dimensional structure with outstanding electrical characteristics such as very high electron mobilities in the order of 20000 $\text{cm}^{2} \text{V}^{-1}\text{s}^{-1}$ on silicon substrates \cite{Chen2008b}⁠. The possibility of achieving such high electron mobilities, which are orders of magnitude higher than silicon based technologies, makes GFETs excellent candidates for replacing nanometer CMOS transistors in future high-speed analog electronic circuits \cite{Schwierz2010a}.

Since the demonstration of the first GFET  \cite{Lemme2007},⁠ the technology has evolved very fast. {In just very few years it has been shown that {de-embedded, intrinsic} GFETs transit frequencies $f_T$ are comparable to or higher than those of similarly sized nanometer CMOS devices \cite{Wu2011a}\cite{Lin2011a}\cite{Liao2010a}⁠. Actual measured $f_T$ is, in fact, much lower than CMOS, mainly due to the presence of interface and contact resistances. These resistances are a serious issue in GFET technology and therefore there are active research efforts on finding ways to reduce  their impact.  Latest research results have shown that contact resistances well bellow 100 $\Omega~\mu$m are possible; for instance, contact resistances as low as 20~$\Omega~\mu$m were measured  for hydrogen intercalated graphene growth \cite{Moon2012}. RF/Analog design uses seldom the minimum width transistors of a technology. Minimum transistor sizes in RF applications are generally  above 20~$\mu$m - 30~$\mu$m, whereas in analog-baseband circuits the dimensions can be as large as hundreds of micrometers. Transistors with these widths would present small contact resistances with values similar to those of parasitic resistances on the metallic interconnection/vias in nanometer CMOS technologies. Their impact on the circuit performance would be the same as other parasitics, and therefore, they can be handled using the same  circuit design techniques that are used in todays CMOS circuits.}

 Likewise, high transconductance gain $g_m$ values were also demonstrated \cite{Moon2010a}⁠\cite{Wu2011a}⁠. In addition, it has been shown that the drain current in GFET transistors has a saturation region \cite{Meric2008}⁠. This is an important characteristic since it facilitates the use the GFETs as voltage-controlled current sources, and consequently, the design of analog circuits in general. {Until now, drain current saturation has been mainly observed in long gate GFET devices, and short-channel GFETs still present unsatisfying current saturation behavior. Nevertheless, it has been reported that the use of bilayer graphene can result  in  important  current saturation improvements \cite{Szafranek2012}. \str{Likewise, lateral graphene heterostructures have also been suggested as a possible solution to enhance the current saturation \cite{Moon2013}}. Although GFET technology still faces technological challenges, projections of GFET vs. CMOS  high-speed analog IC performance \cite{Rodriguez2011} have shown that GFET technology can potentially surpass CMOS in the near future provided that the low field mobility $\mu$ is kept above certain values.} 

The development of GFET devices has been accompanied by the appearance of electrical models that can be used to describe the electrical characteristics of the device and also to simulate circuits \cite{Shepard2008}\cite{Thiele2010}\cite{Jimenez2011a}⁠\cite{Habibpour2011}\cite{Fregonese2012}\cite{Fregonese2013}⁠. Some of these initial models are physical models  which do not have closed expressions and therefore require the use of numerical methods to find solutions. These models are very useful to explain the device physics; however, they are not suitable for implementation in analog circuit modeling languages such as SPICE or Verilog-A. Other models are compact analytical models which can be written in SPICE or Verilog-A and used to simulate circuits with EDA CAD tools.  These models, however, are still very complex  for being used during circuit design. Analog circuit designers make many decisions based on hand-calculations, and therefore require simple analytical expressions.

This paper introduces a comprehensive model which provides the circuit design community with simple mathematical expressions to analyze GFETs. The proposed model is based on⁠ \cite{Fregonese2013}⁠ and consists of simplifications and assumptions which are valid for the first triode region and saturation/negative output resistance region which are relevant for analog circuit design. The paper is organized as follows. Section II presents a brief summary of the large signal model that is used as base of this work. A simplified analytical expression for the drain current as a function of internal voltages and technology parameters is provided in Section III. Section IV provides closed expressions for small-signal hybrid-$\pi$ models ($g_m$, $r_{o}$, $C_{gs}$, $C_{gd}$).  Section V presents closed expression for figures of merit $A_V$, $g_m/id$ and $f_T$ . Finally, a summary of the simplified model is provided. 


\section{Large Signal Model}
An exhaustive study of the drain-source current using the drift equation for GFET transistors can be found in⁠ \cite{Fregonese2013}⁠. The result of this study shows that the drain-source current can be expressed as:
\begin{align}
I_{D} = \mu W\frac{\int_{0}^{V_{DSi}} \left (\left | Q_{net} \right | + en_{puddle}  \right )dV}{L+\mu \left | \int_{0}^{V_{DSi}} \frac{1}{v_{SAT}} dV \right |} \label{EQ:ID_ORIG}
\end{align}
\str{where $\mu$ is the mobility, $W$ the transistor width, $L$ the transistor length,  $Q_{net}$ the net {mobile charge density per unit area, $e$ the elementary charge ($1.6 \times 10^{-19}$ As)}, $ n_{puddle} = \frac{\Delta ^2}{\pi \hbar^2 {v_f}^2 }$, and $V_{DSi}$ the internal drain-source voltage.}  The parameter $\Delta$ represents the spacial inhomogeneity of the electrostatic potential, $\hbar$ is the reduced Planck constant, and $v_f$ is the Fermi velocity. 

For simplicity, the integral in the numerator of  (\ref{EQ:ID_ORIG}) can be split  and solved independently. 
\begin{align}
I_{D} = \mu W\frac{NUM}{DEN}=\mu W\frac{NUM_{1}+ NUM_{2}}{DEN}  \label{EQ:ID_ORIG2}
\end{align}
where the first term in the numerator is:
\begin{equation}
\begin{split}
&NUM_{1}  = {\beta } \int\limits_{0}^{V_{DSi}}  \left[  \frac{-C_{TOP}}{2\beta } + \right.\\
  & \left.  +\frac {\sqrt{{C_{TOP}}^{2} + 4\beta \left |  C_{TOP} (V_{GSi} - V) + eN_{f} \right | }}	{2\beta} \right]^2dV \label{EQ:NUM1_1}
\end{split}
\end{equation}
and the factor $\beta ={e^3}/{(\pi \left ( \hbar  v_f\right ) ^2)}$. $C_{TOP}$ is the top oxide capacitance, $V$ the potential variation along the channel due to $V_{DS}$ and $N_{f}$ is a term that accounts for the net acceptor/donor doping.
Since the graphene material does not have a bandgap, GFETs  do not switch off completely like other FET devices. Instead, they show a minimum conduction point which is known as the Dirac point. The doping level set by  $N_{f}$ is responsible of shifting the Dirac  point in a similar way than the intentional doping used to control the threshold voltage in MOS devices. In practice, the Dirac point is also affected by $V_{DS}$; nevertheless,  $N_{f}$ sets an absolute offset which is biasing independent. Accordingly, it is possible to define a zero-bias threshold voltage for GFET devices as :
\begin{equation}
 V_{TH,0} = eN_{f}/C_{TOP}
\end{equation}
and the effective gate-source overdrive voltage as:
\begin{equation}
V_ {eff} = V_{GSi} + V_{TH,0}
\end{equation}
\str{where $V_{GSi}$ is the internal gate-source voltage. Accordingly,  (\ref{EQ:NUM1_1}) can be rewritten as:}
\begin{equation}
\begin{split}
&NUM_{1}  = {\beta } \int\limits_{0}^{V_{DSi}}  \left[  \frac{-C_{TOP}}{2\beta } + \right.\\
  & \left.  +\frac {\sqrt{{C_{TOP}}^{2} + 4\beta \left |  C_{TOP} (V_{eff} - V) \right | }}	{2\beta} \right]^2dV \label{EQ:NUM1}
\end{split}
\end{equation}
Equation (\ref{EQ:NUM1})  is simplified by introducing the integration variable to $z = C_{TOP}(V_{eff}-V)$. The integral has the following symbolic solution:
\begin{equation}
\begin{split}
&NUM_{1(z>0)}=  - \frac {1}{\beta^{2}{C_{TOP}}}  \left [   \frac{ {C_{TOP}}^{4} }{32} - \right.\\
  &  \left.  \frac {C_{TOP} \left (  {C_{TOP}}^{2} + 4\beta z \right )^{3/2}}{12}+ \frac{\beta^{2}z^{2}}{2} + \frac {\beta {C_{TOP}}^{2}z }{2} \right ]\Bigg|_{z_{1}}^{z_{2}}\\
&NUM_{1(z<0)}=  - \frac {1}{\beta^{2}{C_{TOP}}}  \left [ - \frac{{C_{TOP}}^{4} }{32} + \right.\\
  &  \left.  \frac {C_{TOP} \left (  {C_{TOP}}^{2} - 4\beta z \right )^{3/2}}{12}- \frac{\beta^{2}z^{2}}{2} + \frac {\beta {C_{TOP}}^{2}z }{2} \right ]\Bigg|_{z_{1}}^{z_{2}} \label{EQ:NUM1_2}
\end{split}
\end{equation}
where $z _1= C_{TOP}V_{eff}$ and $z_{2} = C_{TOP}(V_{eff}-V_{DSi})$. 

The second term of the numerator is given by:
\begin{align}
NUM_{2} = \int\limits_{0}^{V_{DSi}} en_{puddle} dV = en_{puddle}V_{DSi}
\end{align}

The denominator in (\ref{EQ:ID_ORIG2}) can be expressed as:
\begin{align}
DEN = L+\mu \left | \int\limits_{0}^{V_{DSi}} \frac{1}{v_{SAT}} dV \right |
\end{align}
which can be simplified assuming an average $v_{SAT}$ given by:
\begin{align}
v_{SAT,AV} = \frac{\omega }{\sqrt{ \pi \frac {\left | Q_{NET,AV} \right | } {e}+n_{puddle}}} \label{EQ:VSATAV}
\end{align}
where $\omega$ is obtained from the surface phonon energy of the substrate $\hbar\omega $ and $Q_{NET,AV}$ is the average charge given by:

\begin{equation}
\begin{split}
& Q_{NET,AV} = \beta \left [ \frac {-C_{TOP}}{2\beta}  + \right.\\
& \left.   
\frac {
\sqrt{ {C_{TOP}}^{2}+4\beta \left | C_{TOP}(V_{eff}-V_{DSi}/2)\right |    } }{2\beta}
\right]^2
\end{split} \label{EQ:QNETAV}
\end{equation}
With the previous assumption, the denominator can be expressed as:
\begin{align}
DEN = L + \frac {\mu}{v_{SAT,AV}} \left | V_{DSi} \right | \label{EQ:DEN}
\end{align}

The accuracy of the model has been successfully evaluated by its authors by comparing it against numeric models and measured data of different GFET devices built by different groups \cite{Meric2010} \cite{Kedzierski2009}⁠ \cite{Moon2010a}⁠.  

The whole model is very compact and perfectly suitable for building SPICE and Verilog-A models; and consequently, it is also suitable for circuit simulation purposes. While the model shows outstanding accuracy, it can be appreciated that it is still complicated to be used by analog circuit designers. The main problem is that it lacks a simple closed mathematical expression for the drain current like the one that is available for CMOS FET transistors (Shichman-Hodges model)  \cite{Harold1968}⁠ or like the collector current in bipolar transistors (Ebers-Moll model) \cite{Ebers1954}⁠. A simple expression for the drain current is fundamental since the parameters for small-signal hybrid-$\pi$ model and figures of merit are directly derived from this equation. The later ones represent the foundation of electronic circuit theory and are the main tools that analog circuit designers have in order to analyze circuit topologies and make design decisions. Therefore, it is of paramount importance to obtain a simplified expression for the GFET drain current. 

\section{Simplified Large Signal Model}

The difficulty in finding a simple expression for the GFET drain current lies in the complexity of (\ref{EQ:NUM1_2}). Fortunately, the replacement of technology dependent parameters taken from the measured GFETs and physical constants unveils that there is a term that dominates and therefore (\ref{EQ:NUM1_2}) can be reduced to:
\begin{align}
NUM_{1} \simeq \left.  - \frac {1}{2} \frac{z^{2}}{ C_{TOP} \times sign(z)} \right |_{z_{1}}^{z_{2}}
\end{align}
which for $z > 0$ (typical case in analog design) can be expressed as:
\begin{align}
NUM_{1} \simeq C_{TOP} V_{DSi} \left (V_{eff}-\frac {V_{DSi}}{2}  \right ) \label{EQ:NUM_SHORT}
\end{align}
For typical technology parameters $NUM_1 \gg NUM_2$. As a result, $NUM_2$ can be disregarded and $NUM \simeq NUM_1$.

{Expression (\ref{EQ:DEN}) becomes complicated when  $v_{SAT,AV}$ is replaced by (\ref{EQ:VSATAV}) and  $Q_{NET,AV}$ by (\ref{EQ:QNETAV}) . }
However, it can also be simplified under the assumption that $V_{eff} > {V_{DSi}}/{2}$, and $n_{puddle} \ll  \pi{ \left | Q_{NET,AV}\right |}/{e}$. Under these conditions, $\left | Q_{NET,AV} \right |\approx C_{TOP}\left ( V_{eff} - V_{DSi} /2\right )$ and the denominator can be simplified to:
\begin{align}
DEN \simeq L + \frac{\mu }{\omega }\sqrt{\frac{\pi C_{TOP}}{e}}V_{DSi}\sqrt{V_{eff}-\frac{V_{DSi}}{2}} \label{EQ:DEN_SHORT}
\end{align}

Finally, the GFET drain current is found by replacing (\ref{EQ:NUM_SHORT}) and (\ref{EQ:DEN_SHORT}) into (\ref{EQ:ID_ORIG2}):
\begin{align}
I_{D} \simeq  
\frac{\mu W C_{TOP} \left ( V_{eff} - {V_{DSi}}/{2} \right )}{\frac{L}{V_{DSi}} + \frac{\mu }{\omega }\sqrt{\frac{\pi C_{TOP}}{e}}\sqrt{V_{eff}-V_{DSi}/2}}\label{EQ:ID_SHORT}
\end{align}

\begin{figure}[t!]
\centering
\includegraphics[width=7.5cm]{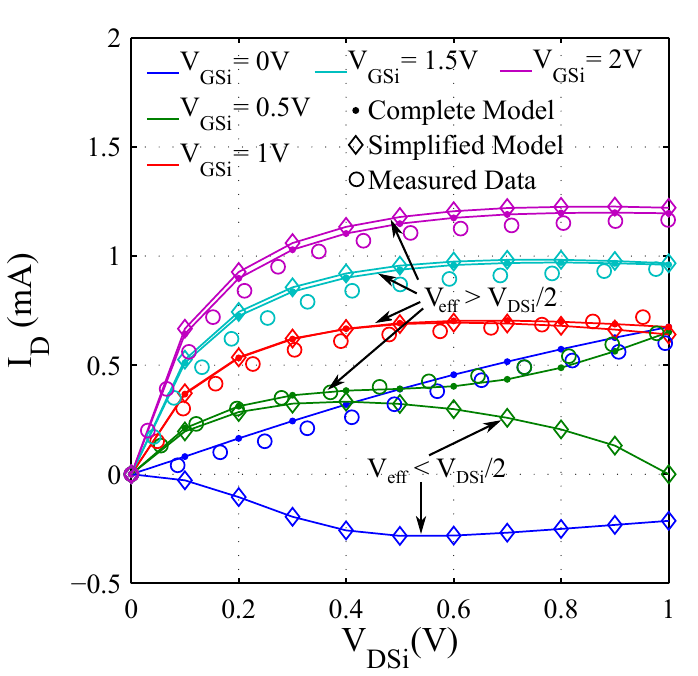}
\caption{Drain current for a 440 nm length, 1 $\mu$m width GFET calculated using the complete model, simplified model, and measured data from \cite{Meric2010}.}
\label{FIG:ID}
\end{figure}
\begin{figure}[t!]
\centering
\includegraphics[width=7.5cm]{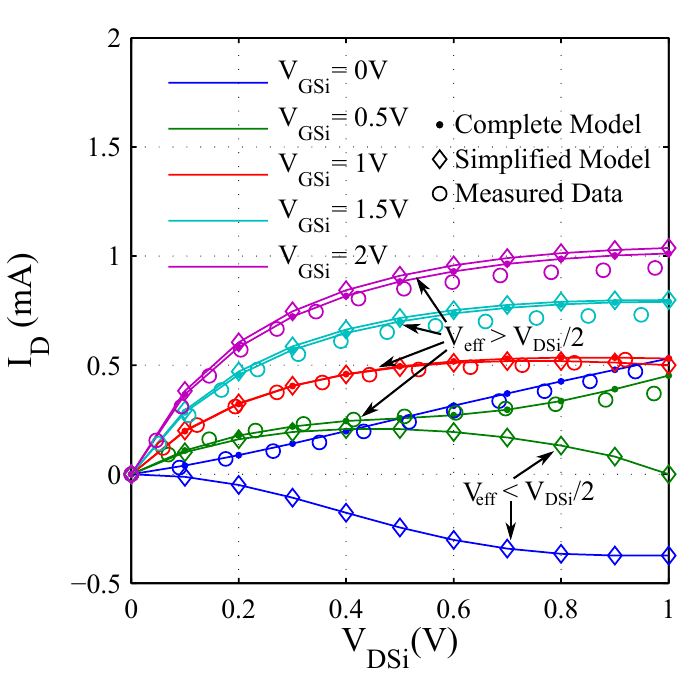}
\caption{Drain current for a 1 $\mu$m length, 1 $\mu$m width GFET calculated using the complete model, simplified model, and measured data from \cite{Meric2010}.}
\label{FIG:ID10}
\end{figure}
\begin{figure}[t!]
\centering
\includegraphics[width=7.5cm]{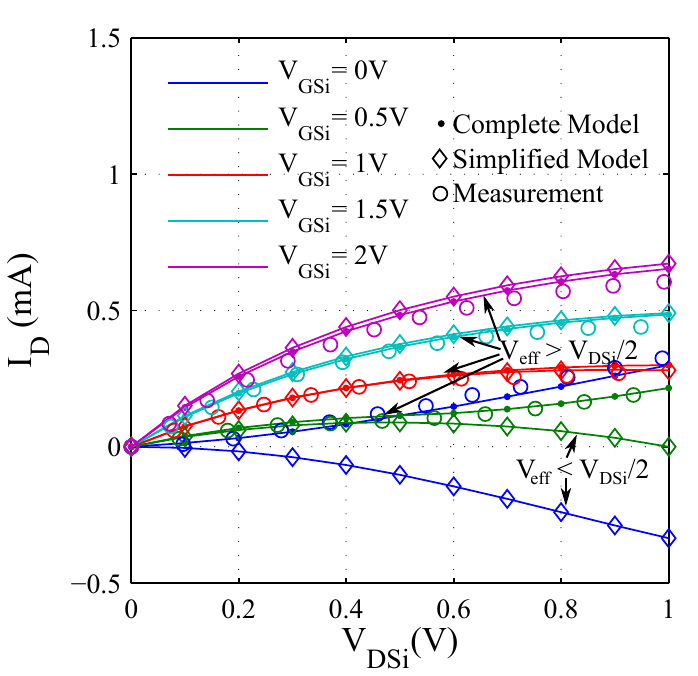}
\caption{Drain current for a 3 $\mu$m length, 1 $\mu$m width GFET calculated using the complete model, simplified model, and measured data from \cite{Meric2010}.}
\label{FIG:ID30}
\end{figure}
which is a closed analytical expression that relates the main technology parameters and biasing conditions. Fig.~\ref{FIG:ID}, Fig.~\ref{FIG:ID10}, and Fig.~\ref{FIG:ID30}  show $I_D$ vs. $V_{DS_i}$ plots for GFETs of 1~um width and 440~nm, 1~um, and 3um length respectively from \cite{Meric2010}. $I_D$ is calculated by using (\ref{EQ:ID_SHORT}) and the complete model including fitting parameters from  \cite{Fregonese2013}.  {For these devices, $N_f$ is approximately 0, and therefore $V_{TH,0} \approx 0$ V. The other parameters have the following values: $C_{TOP} = 3.6 \times 10^{-3} \text{F}/\text{m}^2$, $\mu = 7000$ $\text{cm}^2\text{V}^{-1}\text{s}^{-1}$, and   $\hbar\omega = 56~\text{meV}$.} $V_{GS_i}$ takes values from 0~V to 2~V in steps of 500 mV. It can be appreciated that the simplified model matches very well both the complete model and the measured data for $V_{eff} > V_{DS_i}/2$. The plots show that both the first triode region and the saturation/negative resistance region are correctly modeled by (\ref{EQ:ID_SHORT}). The first triode region can be used to build resistive loads or switches whereas the saturation region can be used to build voltage-controlled current-sources and in some biasing conditions negative resistance loads. The second triode region is not modeled by (\ref{EQ:ID_SHORT}) since the assumption $V_{eff} > V_{DS_i}/2$ does not hold anymore. This region, nevertheless, seems to have little practical value in analog design.

\begin{figure}[t!]
\centering
\includegraphics[width=7.5cm]{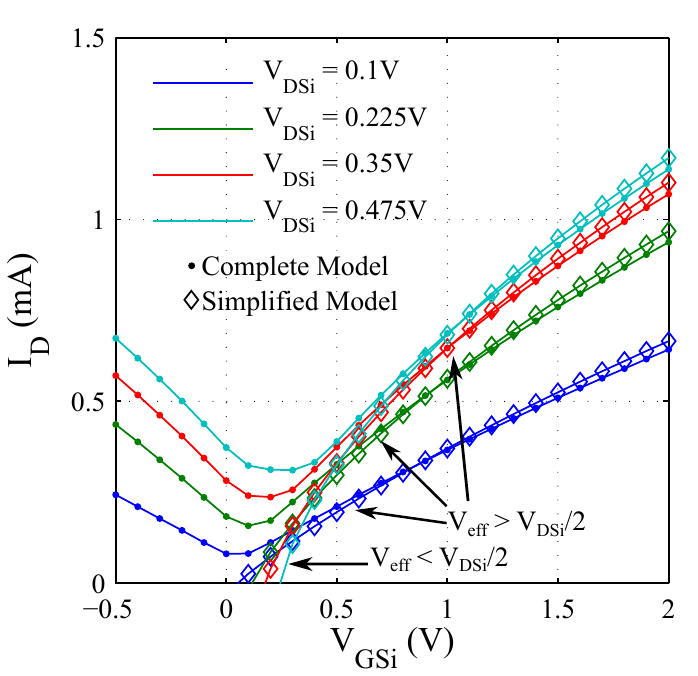}
\caption{Drain current vs. $V_{GSi}$ for a 440 nm length, 1  $\mu$m width GFET from  \cite{Meric2010}  calculated using the complete model, and simplified model}
\label{FIG:ID44_VG}
\end{figure}

{Fig.~\ref{FIG:ID44_VG} shows $I_D$ vs. $V_{GSi}$ curves for the 440~nm lenght, 1~$\mu$m width GFET transistor from \cite{Meric2010}. The ambipolar curves show also that the simplified model follows closely the complete model for $V_{eff} > V_{DS_i}/2$.}

One important observation that can be made from the simplified expressions is the impact of  short-channel lengths on the GFET drain current. It has been experimentally shown that there is strong dependence of short lengths in the GFET transport characteristics \cite{Han2011} \cite{Venugopal2011}⁠. This dependence can be analytically explained by (\ref{EQ:DEN_SHORT}) where it can be seen that   $DEN \simeq L$ only for $V_{DS_i} \approx 0$. Once the gate and drain bias voltages increase, the value of $DEN$ departs from $L$ and increases quickly. 
For very short lengths and high electric fields, the current becomes independent of the channel length, something that has also been confirmed experimentally in \cite{Meric2011}⁠. Under these conditions, the drain current saturates, stops depending on $\mu$, and takes a value of approximately:
\begin{align}
I_{D} \simeq  
 \omega  W  { \sqrt{\frac{ C_{TOP} \times {e} }{\pi}}} \sqrt{V_{eff}-V_{DSi}/2}
 \end{align}


\section{Small Signal Model}

\begin{figure}[t!]
\centering
\includegraphics[width=7.cm]{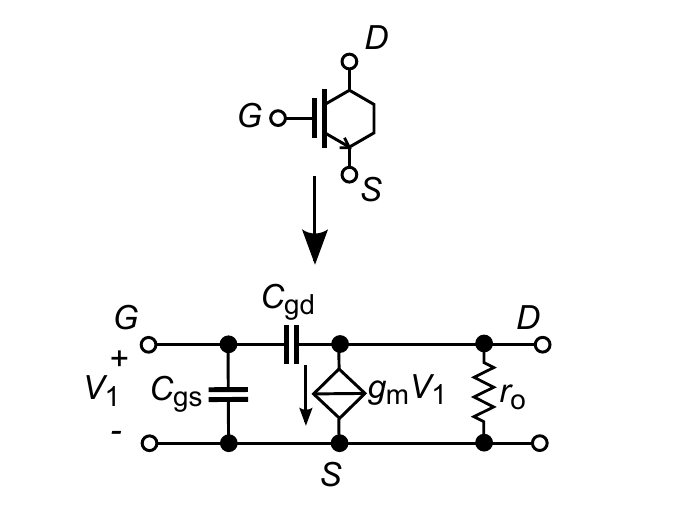}
\caption{GFET Symbol and equivalent hybrid-$\pi$ model for small-signal analysis.}
\label{FIG:SMALLS}
\end{figure}

While the large signal model in (\ref{EQ:ID_SHORT}) encloses the physics of the device in a single expression, it is still too complex to be used in quantitative circuit analyses of the behavior of amplifier configurations.  These analyses are normally performed by taking advantage of linear system theory in which a simplified small-signal representation of the transistor  biased in the operating point is used. The small-signal representation, also called hybrid-$\pi$ model, is shown in Fig. \ref{FIG:SMALLS}. The parameters $g_m$, $r_o$, $C_{gs}$, and $C_{gd}$ can be obtained by linearization of the large signal model. Naturally, the small-signal representation provides only limited information which is valid for small excursions from the operating point. However, it allows to calculate and estimate in an easy way small-signal dynamic linear behavior of gain, phase, poles, zeros, impulse response, etc. Large-signal behavior is non-linear and therefore its analysis requires the use of the complete model and a circuit simulator.  

The derivation of small-signal parameters for the GFET transistor is presented in the following subsections.

\subsection{Transconductance $g_m$}
\begin{figure}[t!]
\centering
\includegraphics[width=7.5cm]{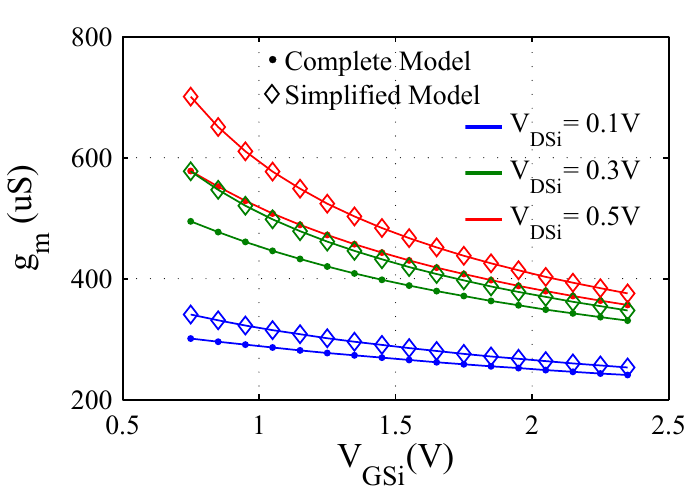}
\caption{Transconductance $g_m$ calculated using the complete and simplified model 
{for the 440 nm length, 1 $\mu$m width GFET from \cite{Meric2010}.  $N_f \approx 0$, $V_{TH,0} \approx 0$ V, $C_{TOP} = 3.6 \times 10^{-3} \text{F}/\text{m}^2$, $\mu = 7000$ $\text{cm}^2\text{V}^{-1}\text{s}^{-1}$, and   $\hbar\omega = 56~\text{meV}$.}
}
\label{FIG:GM}
\end{figure}

The expression for the transconductance gain can be directly derived from  (\ref{EQ:ID_SHORT}):
\begin{align}
\left. g_{m} =\frac{\delta I_{D}}{\delta V_{GSi}}  \right |_{V_{DSi},const.}
\end{align}
\begin{equation}
\begin{split}
&g_{m}=   \left ( \frac{I_{D}}{V_{eff}-V_{DSi}/2} \right ) \left  ( 1  - \frac {1}{2} \frac {I_{D}}{W\omega }\right.\\
  & \left.   \times\sqrt{\frac{\pi }{e \times C_{TOP}}} \frac{1}{ \sqrt{V_{eff} - V_{DSi}/2} }  \right ) \label{EQ:GM}
\end{split}
\end{equation}
Fig. \ref{FIG:GM} shows $g_m$ values calculated using (\ref{EQ:GM}) and the complete model. It can be seen that (\ref{EQ:GM}) follows closely the complete model in particular for $V_{eff} > V_{DSi}/2 $.
{It is interesting to notice that $g_{m}$ drops substantially at large $V_{GSi}$ biasing voltages, mainly due to the effect of $V_{SAT}$. Therefore,  the best $g_{m}$ performance is actually achieved at low $V_{eff}$ voltages.}

\subsection{Output resistance $r_o$}
The output resistance can be calculated as $r_{0} = 1/g_{0}$ where $g_{0}$ is the output conductance. An expression for $g_o$ can also be directly derived from (\ref{EQ:ID_SHORT}):
\begin{align}
g_{o} =  \left. \frac{\delta I_{D}}{\delta V_{DSi}} \right |_{V_{GSi,const.}}
\end{align}
\begin{align}
\begin{split}
&g_{o} = \frac{I_{D}}{V_{eff}-V_{DSi}/2}\left [ -\frac{1}{2} + \right.\\
 & +\left. \frac{I_{D}}{\mu WC_{TOP}
} \left (  \frac{L}{{V_{DSi}}^{2}} + \frac{ \frac{\mu }{\omega }\sqrt{\frac{\pi C_{TOP}}{e}} }{4 \sqrt{V_{eff}-V_{DSi}/2}}\right )\right ] \label{EQ:GO}
\end{split}
\end{align}

One important characteristic of the GFET device is that under some biasing conditions, $g_0$ becomes negative \cite{Wu2012}⁠ \cite{Grassi2013}⁠. A negative $g_0$ makes the  device unstable, and in general this region needs to be avoided  in amplifier design. On the other hand, a negative $g_0$ is a very welcome asset when designing oscillators. The biasing conditions in which $g_o$ changes from positive to negative values can be found by making $g_0 = 0$ in (\ref{EQ:GO})  and solving for $V_{DS}$. The expression for this boundary condition is:
\begin{align}
V_{DS,lim} = \frac{-2L+\sqrt{ 4L \left (L + \frac{\mu}{\omega } \sqrt{\frac{\pi C_{TOP}}{e}} {V_{eff}}^{3/2}  \right )   }}
{  
\frac{\mu}{\omega } \sqrt{\frac{\pi C_{TOP}}{e} }\sqrt{V_{eff}}
} \label{EQ:RES}
\end{align}

\begin{figure}[t!]
\centering
\includegraphics[width=7.5cm]{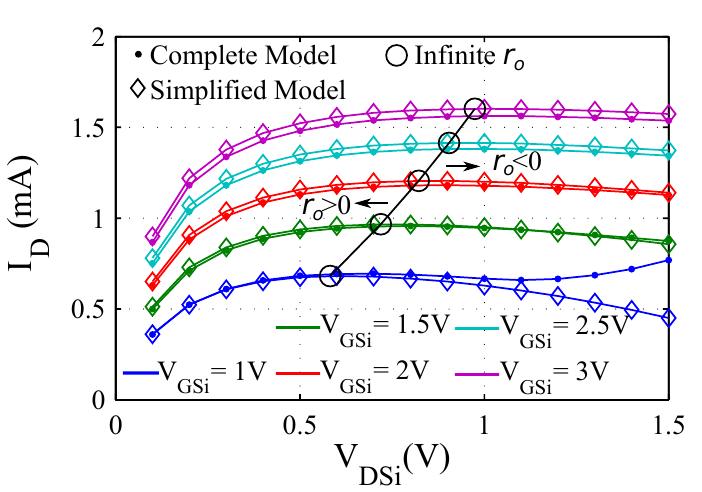}
\caption{ Calculation of negative $r_o$ biasing requirements 
{for the 440 nm length, 1 $\mu$m width GFET from \cite{Meric2010}.  $N_f \approx 0$, $V_{TH,0} \approx 0$ V, $C_{TOP} = 3.6 \times 10^{-3} \text{F}/\text{m}^2$, $\mu = 7000$ $\text{cm}^2\text{V}^{-1}\text{s}^{-1}$, and   $\hbar\omega = 56~\text{meV}$.}
}
\label{FIG:RES}
\end{figure}

Fig \ref{FIG:RES} shows $I_D$ vs.  $V_{DS}$ plots for different $V_{GS}$ voltages. In addition, the plot shows the points in which $g_o = 0$  ($r_o = \infty$) which were found by using (\ref{EQ:RES}). It can be seen that (\ref{EQ:RES}) predicts very well the transition from positive to negative output resistance.

\subsection{Total Channel Charge}
The distributed gate capacitance in GFET devices is modeled as the series capacitance of $C_{TOP}$ and the quantum capacitance $C_q$ in the graphene channel \cite{Thiele2010}⁠: 
\begin{align}
Cg = \frac{C_{q} \times C_{TOP}}{C_{q} + C_{TOP}}
 \end{align}
where the parameter $C_q$ relates the distributed charge along the graphene channel and its potential $V_{CH}$, which depends strongly on both $V_{GS}$ and $V_{DS}$. 
{The total charge stored in the gate capacitance can be found by considering that the charge of all capacitors is the same when they are connected in series. Accordingly, the total charge stored in the gate capacitance is equal to the total charge in the graphene channel $Q_{CH}$}.
The separation of the gate capacitance between Gate-Source capacitance and Gate-Drain capacitance can be done by taking partial derivatives of $Q_{CH}$. Consequently, it is beneficial to find a simple closed expression for $Q_{CH}$ which can be easily differentiated. $Q_{CH}$ can be expressed as \cite{Fregonese2013}⁠:
\begin{align}
Q_{CH} = W\int\limits_{0}^{L} \left (Q_{NET}(x)+en_{puddle}  \right )dx \label{EQ:QCH}
\end{align}
By changing the integration variable dx to dV and reordering the expression, $Q_{CH}$ becomes: 
\begin{align}
Q_{CH} = \frac{eW}{E_{AV}} \int\limits_{0}^{V_{DSi}}\left ( \frac{\beta }{e} \left | V_{CH} \right | V_{CH} + n_{puddle} \right )dV \label{EQ:QCH2}
\end{align}
where $E_{AV}$ is the average electric field which is given by:
\begin{align}
E_{AV} \approx \frac{dV}{dx} \approx \frac{V_{DSi}}{L} \label{EQ:EAV_APROX}
\end{align}

The integral in (\ref{EQ:QCH2})  is similar to that in (\ref{EQ:NUM1}) and therefore it is solved in the same way. Likewise, there is a quadratic term that dominates and therefore $Q_{CH}$ can be reduced to:
\begin{align}
Q_{CH} \approx \left. \frac{e \times W}{2E_{AV}}\left (-\frac{z^2}{C_{TOP} \times e} \right ) \right |_{z_{1}}^{z_{2}} 
\end{align}
which after replacing $z_1$ and  $z_2$ becomes:
\begin{align}
Q_{CH} \approx \frac{W C_{TOP}}{ E_{AV}} V_{DSi}\left ( V_{eff} - \frac{V_{DSi}}{2}\right ) \label{EQ:QCH3}
\end{align}

Finally, a simplified expression for $Q_{CH}$ is found by replacing (\ref{EQ:EAV_APROX}) into (\ref{EQ:QCH3}):
\begin{align}
Q_{CH} \approx C_{TOP}WL \left (   V_{eff} - V_{DSi}/2\right) \label{EQ:QCH_SHORT}
\end{align}



\subsection{Gate-Source Capacitance $C_{gs}$}
The small-signal gate-source capacitance can be calculated as:
\begin{align}
C_{gs} = \left. \frac{\delta Q_{CH}}{\delta V_{GSi}} \right |_{V_{DSi,const.}}
\end{align}
\begin{align}
C_{gs} =C_{TOP}WL \label{EQ:CGS}
\end{align}

\begin{figure}[t!]
\centering
\includegraphics[width=7.5cm]{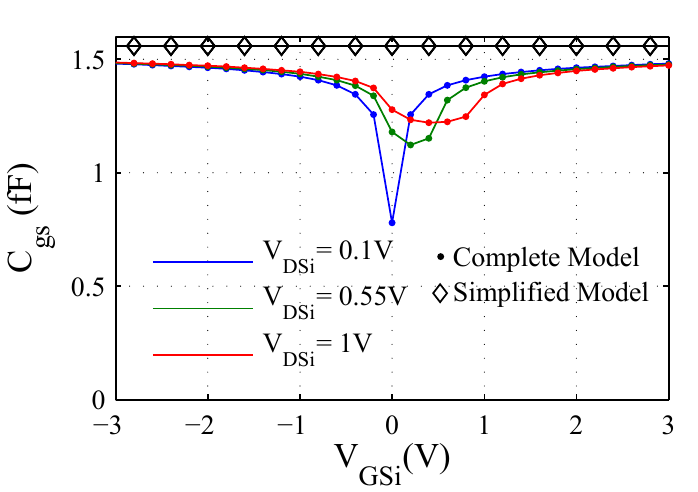}
\caption{ $C_{gs}$ calculation using the complete and simplified model
{for the 440 nm length, 1 $\mu$m width GFET from \cite{Meric2010}.  $N_f \approx 0$, $V_{TH,0} \approx 0$ V, $C_{TOP} = 3.6 \times 10^{-3} \text{F}/\text{m}^2$, $\mu = 7000$ $\text{cm}^2\text{V}^{-1}\text{s}^{-1}$, and   $\hbar\omega = 56~\text{meV}$.}
}
\label{FIG:CGS}
\end{figure}

Fig. \ref{FIG:CGS} shows plots of $C_{gs}$ calculated using (\ref{EQ:CGS}) and the complete model. It can be seen that for $V_{eff} > V_{DS}/2$, $C_{gs}$ approaches the value of the total oxide capacitance. For large $V_{eff}$ values the error is within 5\%. 

\subsection{Gate-Drain Capacitance $C_{gd}$}
The small-signal gate-drain capacitance can be calculated as:

\begin{align}
C_{gd} = -\left. \frac{\delta Q_{CH}}{\delta V_{DSi}} \right |_{V_{GSi,const.}}
\end{align}
\begin{align}
C_{gd}= \frac {C_{TOP}WL}{2} \label{EQ:CGD}
\end{align}

\begin{figure}[t!]
\centering
\includegraphics[width=7.5cm]{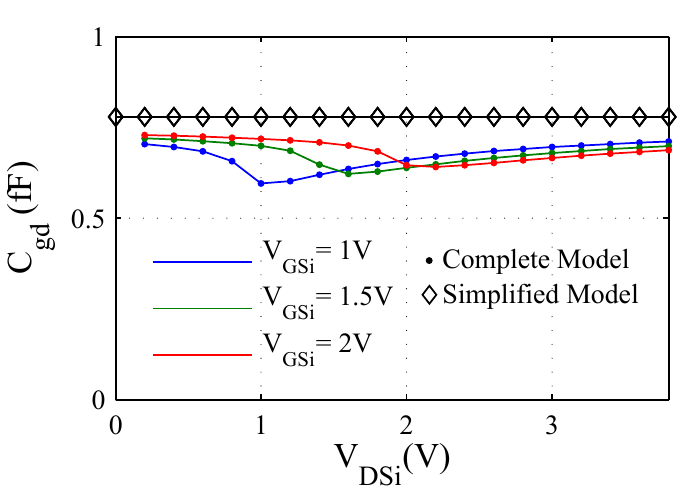}
\caption{ $C_{gd}$ calculation using the complete and simplified model
{for the 440 nm length, 1 $\mu$m width GFET from \cite{Meric2010}.  $N_f \approx 0$, $V_{TH,0} \approx 0$ V, $C_{TOP} = 3.6 \times 10^{-3} \text{F}/\text{m}^2$, $\mu = 7000$ $\text{cm}^2\text{V}^{-1}\text{s}^{-1}$, and   $\hbar\omega = 56~\text{meV}$.}
}
\label{FIG:CGD}
\end{figure}
Fig. \ref{FIG:CGD} shows plots of $C_{gd}$ calculated using  (\ref{EQ:CGD}) and the complete model. In this case it is also possible to see that even though $C_{gd}$ values calculated with the complete model have valleys at different drain biasing conditions, their values are close to the value predicted by (\ref{EQ:CGD}). For large $V_{DS}$ values the error is within 15\%. 

\begin{figure}[t!]
\centering
\includegraphics[width=8.5cm]{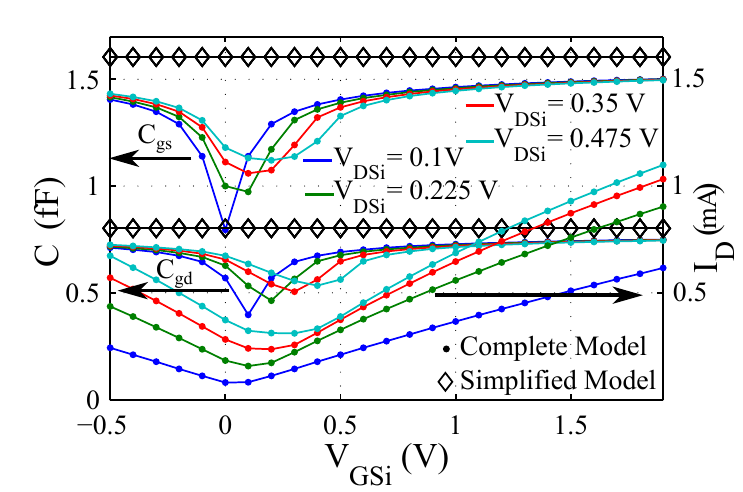}
\caption{
{ $C_{gs}$, $C_{gd}$, and $I_D$ calculation using the complete and simplified model
for the 440 nm length, 1 $\mu$m width GFET from \cite{Meric2010}.  $N_f \approx 0$, $V_{TH,0} \approx 0$ V, $C_{TOP} = 3.6 \times 10^{-3} \text{F}/\text{m}^2$, $\mu = 7000$ $\text{cm}^2\text{V}^{-1}\text{s}^{-1}$, and   $\hbar\omega = 56~\text{meV}$.}
}
\label{FIG:CGS_CGD}
\end{figure}

{Fig. \ref{FIG:CGS_CGD} shows $C_{gs}$, $C_{gd}$, and $I_{D}$ vs. $V_{GSi}$. It is interesting to see that despite the fact that the capacitances change for different biasing conditions, their values approximate very well the simplified model when enough $V_{GSi}$ and $V_{DSi}$ bias is present.}

\section{Figures of Merit}
The extraction of small-signal parameters allows the calculation of figures of merit that can be used to make performance comparisons. The main figures of merit used to evaluate amplifying devices are: intrinsic voltage gain $A_V$, transconductance efficiency $g_m/I_D$, and transit frequency $f_T$. Expressions for these figures of merit are found in the following subsections.
\subsection{Intrinsic Voltage Gain $A_V$}
The intrinsic voltage gain estimates the low frequency voltage amplification capabilities of the device and can be calculated as:
\begin{align}
A_V = g_m \times r_0
\end{align}
\begin{equation}
\begin{split}
A_{V} &=  \left(	{1 - \frac{I_{DSi}}{2W\omega }  \sqrt{\frac{\pi }{e C_{TOP}}} \frac{1}{\sqrt{V_{eff} - V_{DSi}/2}}} \right)\bigg/   \\
  & \left[ -\frac{1}{2} +   \frac{I_{DSi}}{\mu  W  C_{TOP}}\left(  \frac{L}{{V_{DSi}}^{2}} +\frac{\mu }{4 \omega } \sqrt{\frac{\pi C_{TOP}}{e}}\times   \right.   \right.\\
  & \left. \left.    \frac{1}{\sqrt{V_{eff} - V_{DSi}/2}}\right) \right]\label{EQ:GAIN}
\end{split}
\end{equation}
\subsection{Transconductance Efficiency $g_m/I_D$}

The $g_m/I_D$ relates the transconductance amplification capability of the device and the     drain current that is required to produce it. Therefore, it is a measure of the power consumption efficiency of the amplifying device.  
The expression for $g_m/I_D$ can be directly obtained from (\ref{EQ:GM}):
\begin{equation}
\begin{split}
\frac { g_{m}} {I_{D}} &= \left ( \frac{1}{V_{eff}-V_{DSi}/2} \right ) \times\\
&\left ( 1 - \frac {1}{2} \frac {I_{D}}{W\omega }
\sqrt{\frac{\pi }{e \times C_{TOP}}}\frac{1}{ \sqrt{V_{eff} - V_{DSi}/2} }
\right )
\label{EQ:GMID}
\end{split}
\end{equation}



\subsection{Transit Frequency $f_T$}
The transit frequency estimates the frequency at which the current gain of the device drops to 1, and it is a measure of its high-speed and bandwidth capabilities. The transit frequency is defined as:
\begin{align}
f_{T} = \frac {g_{m}}{2  \pi \left (  C_{gs}  + C_{gd} \right ) }
\end{align}
\begin{equation}
\begin{split}
f_{T} &= \frac{\left ( \frac{I_{D}}{V_{eff}-V_{DSi}/2} \right )}{2\pi \left (\frac {3}{2} C_{TOP} W L  \right )}\times \\
 &\left ( 1 - \frac {1}{2} \frac {I_{D}}{W\omega }
\sqrt{\frac{\pi }{e \times C_{TOP}}}\frac{1}{ \sqrt{V_{eff} - V_{DSi}/2} }
\right )
\label{EQ:FT}
\end{split}
\end{equation}

\begin{figure}[t!]
\centering
\includegraphics[width=7.5cm]{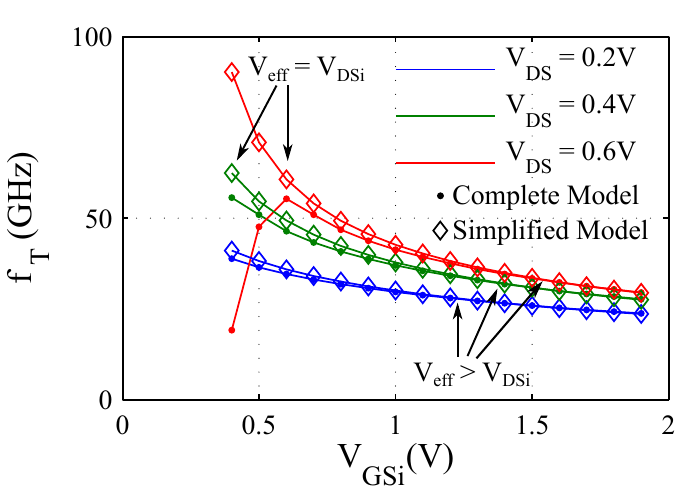}
\caption{ Transit Frequency $f_T$ calculation using the complete and simplified model
{for the 440 nm length, 1 $\mu$m width GFET from \cite{Meric2010}.  $N_f \approx 0$, $V_{TH,0} \approx 0$ V, $C_{TOP} = 3.6 \times 10^{-3} \text{F}/\text{m}^2$, $\mu = 7000$ $\text{cm}^2\text{V}^{-1}\text{s}^{-1}$, and   $\hbar\omega = 56~\text{meV}$.}
}
\label{FIG:FT}
\end{figure}

Fig. \ref{FIG:FT} shows plots of $f_T$ calculated using (\ref{EQ:FT}) and the complete model. It can be seen that there is very good  matching for most biasing points, and disagreements start to become visible only when $V_{eff} < V_{DS}.$ 

\section{Summary}

\begin{table}[!t]
\renewcommand{\arraystretch}{1.5}
\caption{Summary of the Simplified GFET Model }
\centering
\begin{tabular}{c c c}
\hline
\bfseries Name & \bfseries Expression  & \bfseries Units \\
\hline\hline
$I_D$ & $\frac{\mu W C_{TOP} \left ( V_{eff} - {V_{DSi}}/{2} \right )}{{L}/{V_{DSi}} + \frac{\mu }{\omega }\sqrt{{\pi C_{TOP}}/{e}}\sqrt{V_{eff}-V_{DSi}/2}}$ & [A]\\
$V_ {eff}$  & $V_{GSi} +  eN_{f}/C_{TOP}$ & [V] \\
$C_{gs}$ & $C_{TOP}WL$ & [F]\\
$C_{gd}$ & ${C_{TOP}WL}/{2}$ & [F] \\
\hline
\end{tabular}
\label{TABLE:PAR}

\end{table}
A summary of the simplified GFET model is shown in Table~\ref{TABLE:PAR}. The expressions in this table were used to extract small-signal hybrid-$\pi$ model parameters and figures of merit typically used to compare the performance of transistors. 
The proposed model has been validated by comparing it against a complete analytical model and to measured data available in current literature. Whereas the complete analytical model hides the effects of physical parameters behind many separate calculations, the proposed model provides a simple expression that enables direct identification of dominant physical parameters. In addition, the proposed GFET model is ready for use in circuit design in exactly  the same way as the Shichman-Hodges and Ebers-Moll models are used for CMOS and bipolar circuit design respectively.

\ifCLASSOPTIONcaptionsoff
  \newpage
\fi



\bibliographystyle{IEEEtran}
\bibliography{GRAPHENE_CIRCUITS}
%
%

%


\begin{IEEEbiography}[{\includegraphics[width=1in,height=1.25in,clip,keepaspectratio]{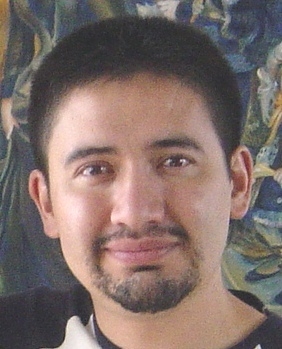}}]%
{Saul Rodriguez}
received the B.Sc. degree in Electrical Engineering from the Army Polytechnic School (ESPE), Quito, Ecuador in 2001, the M.Sc. degree in System-on-Chip Design in 2005 and the Ph.D. degree in Electronic and Computer Systems in 2009 from the Royal Institute of Technology (KTH), Stockholm, Sweden. His research area covers from RF CMOS circuit design for wideband front-ends, ultra-low power circuits for medical applications, and graphene-based RF and AMS circuits.
\end{IEEEbiography}

\begin{IEEEbiography}[{\includegraphics[width=1in,height=1.25in,clip,keepaspectratio]{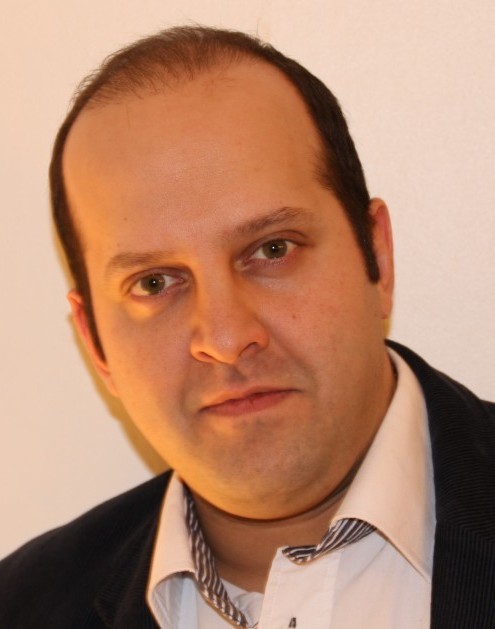}}]%
{Sam Vaziri}
received the B.Sc. and  M.Sc. in Solid State Physics from Shahid Beheshti University and K.N.Toosi University of Technology, Tehran, Iran. In 2011, he received his second M.Sc. degree in Nanotechnology from KTH Royal Institute of Technology, Stockholm, Sweden. He is currently a PhD candidate at KTH, investigating graphene electronic and optoelectronic devices.
\end{IEEEbiography}

\begin{IEEEbiography}[{\includegraphics[width=1in,height=1.25in,clip,keepaspectratio]{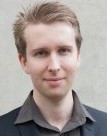}}]%
{Anderson Smith}
received his bachelors and masters degrees from the Georgia Institute of Technology in mechanical engineering and then a second masters degree in nanotechnology at KTH Royal Institute of technology.  He is currently working on his Ph.D. at KTH Royal Institute of Technology in the school of Information and Communication Technology.  His research focuses primarily on graphene sensors and transistors for more and Moore applications.
\end{IEEEbiography}

\begin{IEEEbiography}[{\includegraphics[width=1in,height=1.25in,clip,keepaspectratio]{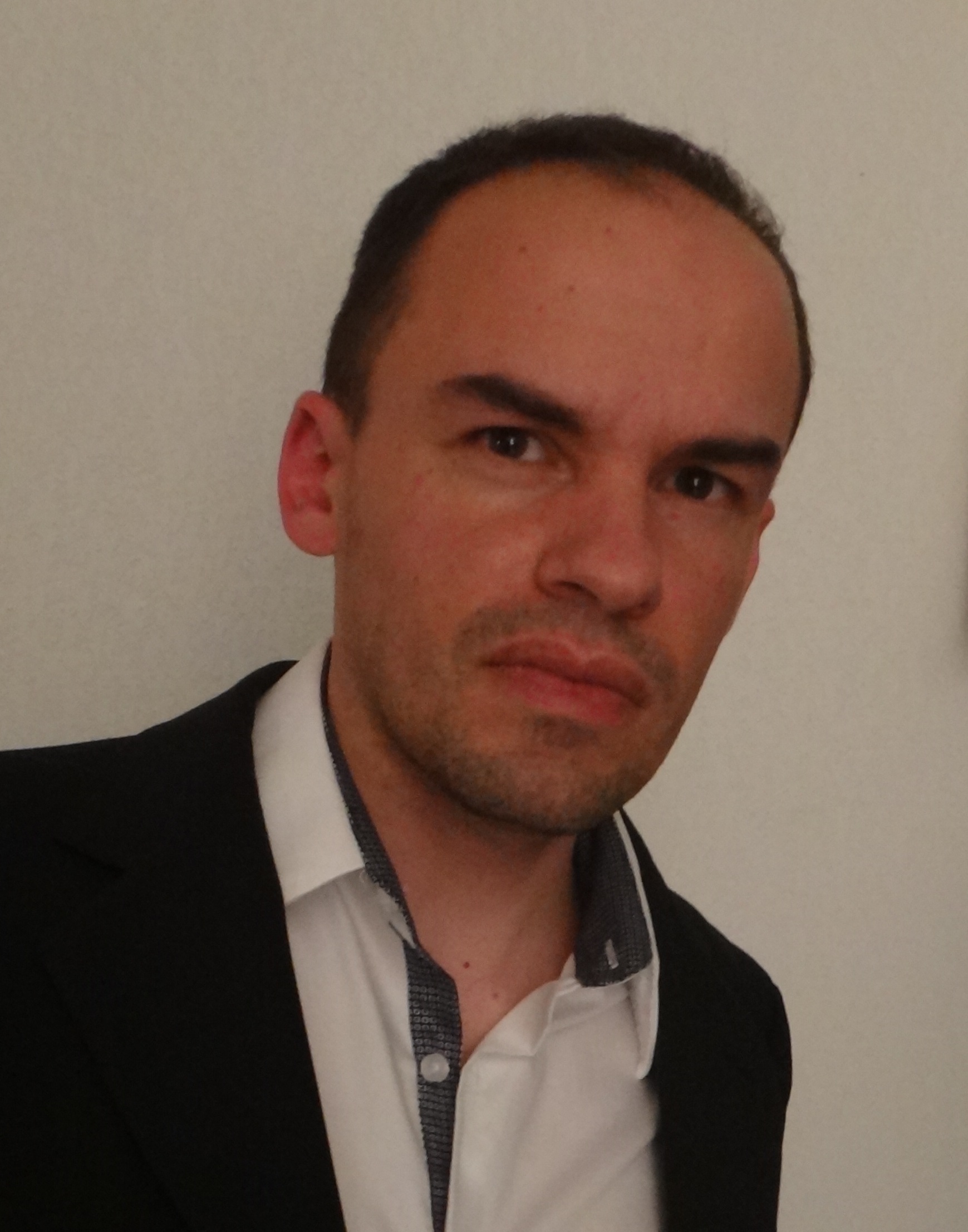}}]%
{S\'ebastien~Fr\'egon\`ese}
received the M.Sc. and the Ph.D. degrees in electronics from the Universit Bordeaux, Talence, France, in 2002 and 2005, respectively. From 2005 to 2006, he was with the Technical University of Delft, The Netherlands. In 2007, he joined the Centre
National de la Recherche Scientifique (CNRS), France. His research interests focus on electrical compact modeling and characterization of devices such as the SiGe HBTs and carbon based transistors.
\end{IEEEbiography}

\begin{IEEEbiography}[{\includegraphics[width=1in,height=1.25in,clip,keepaspectratio]{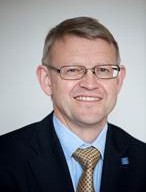}}]%
{Mikael Ostling}
(M’85–SM’97–F’04) received the Ph.D. degree from Uppsala University, Uppsala, Sweden, in 1983. He is a Professor and Head of the Department of Integrated Devices and Circuits with the School of Information and Communication Technology, KTH.
\end{IEEEbiography}






\begin{IEEEbiography}[{\includegraphics[width=1in,height=1.25in,clip,keepaspectratio]{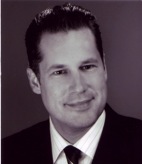}}]%
{Max Lemme} received the Dr.-Ing. (2003) degree in Electrical Engineering from RWTH Aachen University, Germany. He is Heisenberg-Professor for Graphene-based Nanotechnology at the University of Siegen, Germany. From 2010-2013 He was Guest Professor at KTH, Sweden and from 2008-2010 he was a research fellow at Harvard University. From 1998-2008 he worked at nanotechnology start-up AMO GmbH, Germany, as Head of Technology Department. His research interests include non-conventional nano-CMOS devices and graphene and 2D-material technology, devices and circuits.
\end{IEEEbiography}

\begin{IEEEbiography}[{\includegraphics[width=1in,height=1.25in,clip,keepaspectratio]{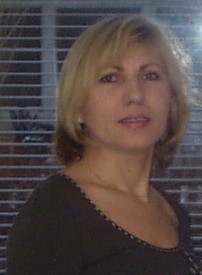}}]%
{Ana Rusu}

received degrees of MSc (1983) in Electronics and Telecommunications and PhD (1998) in Electronics. Since September 2001, she has been with KTH Royal Institute of Technology, Stockholm, Sweden, where she is Professor in Electronics Circuits for Integrated Systems. Her research interests spans from low/ultra-low power high performance CMOS circuits and systems for a wide range of applications to circuits using emerging technologies, such as graphene and SiC. 
\end{IEEEbiography}


\vfill


\end{document}